\documentclass[11pt,american,twoside,a4wide]{article}
\usepackage[T1]{fontenc}
\usepackage[latin1]{inputenc}
\usepackage{latexsym}
\usepackage{amsmath}
\usepackage{bbm}
\usepackage{amssymb}
\usepackage{babel}
\usepackage{amsfonts}
\usepackage{times}
\usepackage{theorem}
\usepackage{epsfig}
\usepackage{enumerate}
\usepackage{color}
\usepackage[active]{srcltx}
\usepackage[colorlinks=true]{hyperref}
\usepackage{fancyhdr}

\newcounter{smallarabics}
\newenvironment{arabicenumerate}
{\begin{list}{{\normalfont\textrm{(\arabic{smallarabics})}}}
  {\usecounter{smallarabics}\setlength{\itemindent}{0cm}
   \setlength{\leftmargin}{5ex}\setlength{\labelwidth}{4ex}
   \setlength{\topsep}{0.75\parsep}\setlength{\partopsep}{0ex}
   \setlength{\itemsep}{0ex}}}
{\end{list}}

\newcounter{smallroman}

\newcommand{\ben}{\begin{arabicenumerate}}  
\newcommand{\een}{\end{arabicenumerate}}  

\newtheorem{theoreme}{Theorem}[section]
\newtheorem{proposition}[theoreme]{Proposition}
\newtheorem{lemma}[theoreme]{Lemma}
\newtheorem{definition}[theoreme]{Definition}

\def\textsl{{}}

\def\c0inf{C_0^\infty}
\def\bep{\begin{proposition}}
\def\eep{\end{proposition}}

\def\i{{\rm i}}
\newcommand{\beq}{\begin{equation}}
\newcommand{\eeq}{\end{equation}}
\newcommand{\bear}[1]{\begin{array}{#1}}
\newcommand{\ear}{\end{array}}

\newcommand{\e}{\mathrm{e}}
\renewcommand{\i}{\mathrm{i}}

\def\bel{\begin{lemma}}
\def\eel{\end{lemma}}
\def\bet{\begin{theoreme}}
\def\eet{\end{theoreme}}
\def\bed{\begin{definition}}
\def\eed{\end{definition}}

\def\12{\frac{1}{2}}

\def\e{{\rm e}}

\def\tr{{\rm tr}}

\begin{document}
\def\today{}
\title{Entropic Functionals  in Quantum Statistical Mechanics}
\author{Vojkan  Jak\v{s}i\'c$^{1}$ and  Claude-Alain  Pillet$^2$
\\ \\
$^1$Department of Mathematics and Statistics\\ 
McGill University\\
805 Sherbrooke Street West \\
Montreal,  QC,  H3A 2K6, Canada
\\ \\
$^2$Aix-Marseille Universit\'e, CNRS UMR 7332, CPT, 13288 Marseille, France\\
Universit\'e du Sud Toulon-Var, CNRS UMR 7332, CPT, 83957 La Garde, France
}
\maketitle
\thispagestyle{empty}

\begin{quote}
\centerline{To appear in:}
\centerline {\em Proceedings of the XVIIth International Congress of Mathematical Physics, Aalborg, Denmark, 2012}
\end{quote}
\begin{quote}
\noindent{\bf Abstract.}
We describe  quantum  entropic functionals  and  outline a research program dealing 
with entropic fluctuations in non-equilibrium quantum statistical mechanics.
\end{quote}
\section{Introduction}
Starting with  papers \cite{JP1,JP2,Ru1,Ru2}, the mathematical theory of non-equilibrium quantum statistical mechanics has developed rapidly in the last decade. 
The initial developments concerned the theory of  non-equilibrium steady states, the entropy production observable, and linear response theory 
(Green-Kubo formulas, Onsager reciprocity relations) 
for open systems driven by thermodynamical forces (say temperature  differentials). This line of development was a  
natural direct quantization of the classical theory.  In contrast,  extensions  of the classical 
fluctuation relations of Evans-Searles \cite{ES} and Gallavotti-Cohen \cite{GC} to the quantum domain have led to some  surprises and  novel classes of 
entropic functionals with somewhat striking mathematical structure and physical interpretation. 

A pedagogical introduction to our  research  program dealing with fluctuation theorems/relations in  non-equilibrium quantum statistical mechanics 
 can be found in  \cite{JOPP}.  This note can be viewed 
as a brief introduction to \cite{JOPP}. We sketch in telegraphic and simple  terms the finite time/finite volume theory in the  classical case 
(for comparison purposes) and in the quantum case, and comment on the resulting research program. The interested reader may consult 
\cite{JPR, JOPP}  for additional information.

\section{Classical picture}
Consider  a pair $({\cal E}, \phi)$, where ${\cal E}=\{ \zeta_j\}_{0\leq j\leq N}$ is a
finite phase space  and $\phi: {\cal E}\rightarrow {\cal E}$ is a discrete time  dynamics. For our purposes 
without loss of generality we may assume that $\phi(\zeta_j)=\zeta_{j+1}$ ($\zeta_{N+1}=\zeta_0$).  
Observables are functions $f: {\cal E}\rightarrow {\mathbb R}$ and  states 
are non-vanishing probability measures $\rho$ on ${\cal E}$. We write 
$\rho(f)=\sum_\zeta f(\zeta)\rho(\zeta)$. Observables evolve in time as $f_t =f \circ\phi^t$, $t\in {\mathbb Z}$, and states as $\rho_t(f)=\rho(f_t)$. 

The relative entropy of two states
\[
S(\rho, \nu)=\sum_\zeta \rho(\zeta)\log\left(\nu(\zeta)/\rho(\zeta)\right),
\] 
satisfies $S(\rho, \nu)\leq 0$ and  $S(\rho, \nu)=0$ iff $\rho=\nu$. 
The R\'enyi relative entropy of order $\alpha \in {\mathbb R}$ is defined by
\[S_\alpha(\rho, \nu)=\log \sum_\zeta \rho(\zeta)^{1-\alpha}\nu(\zeta)^\alpha.\]

Our  starting point is a dynamical system $({\cal E}, \phi, \omega_0)$ where $\omega_0$ is a given reference state. We assume that 
$\omega_0$ is not constant (and hence  not invariant under $\phi$) and that $\omega_0(\zeta_j) =\omega_0(\zeta_{N-j})$. This last assumption 
ensures that our dynamical system is  time reversal invariant (TRI) with time reversal $\theta(\zeta_j)=\zeta_{N-j}$. 
 
The entropy observable is ${\cal S}_0(\zeta)=-\log\omega_0(\zeta)$. The observable $\Sigma^t = ({\cal S}_t- {\cal S}_0)/t$ describes the mean entropy production rate over the time interval $[0,t]$.
One easily verifies that for $t>0$,
\begin{equation}
\omega_0(\Sigma^t)= -\frac{1}{t}S(\omega_t, \omega_0)\geq 0,
\label{sec-law-cl}
\end{equation}
in accordance with the (finite time) second law of thermodynamics. The entropy production observable (or the phase space contraction rate) is defined by $\sigma(\zeta)=-\log\left(\omega_1(\zeta)/\omega_{0}(\zeta)\right)$ and satisfies
$\Sigma^t=t^{-1}\sum_{s=1}^{t}\sigma_s$.

The relation (\ref{sec-law-cl}) holds without the TRI assumption.  The TRI however allows to refine the  second law as follows. Let 
 ${\cal E}_{ t\lambda}=\{ \zeta\,|\, \Sigma^t(\zeta)=\lambda\}$ and $p^t(\lambda)= \omega_0({\cal E}_{ t\lambda})$.  An easy computation gives the celebrated Evans-Searles 
 fluctuation relation
\begin{equation}
p^t(-\lambda) ={\rm e}^{-\lambda t}p^t(\lambda).
\label{es-rel}
\end{equation}
This relation implies $\omega_0(\Sigma^t)\geq 0$ and is saying more: the negative values of the mean entropy 
production rate are exponentially suppressed  in a universal manner.

The classical entropic functional is defined by 
\begin{equation} e_t(\alpha)=\log \omega_0\left({\rm e}^{-\alpha t\Sigma^t}\right).
\label{cl-ef}
\end{equation}
The symmetry 
\begin{equation}
e_t(\alpha)=e_t(1-\alpha),
\label{es-rel1}
\end{equation}
which holds for all $\alpha \in {\mathbb R}$, is an equivalent formulation of the fluctuation relation (\ref{es-rel}). Clearly, $e_t(0)=0$, and hence $e_t(1)=0$ (this is sometimes called 
the Kawasaki identity \cite{CWWSE}). The function 
$\alpha \mapsto e_t(\alpha)$ is convex, and $e_t^\prime(0)=-t \omega_0(\Sigma^t)$.  The classical entropic functional satisfies
\begin{equation}
e_t(\alpha)=\max_\rho S(\rho, \omega_0)-\alpha t \rho(\Sigma^t),
\label{var}
\end{equation}
and 
\begin{equation}
e_t(\alpha)=S_\alpha(\omega_t, \omega_0). 
\label{cl-re}
\end{equation}
The functional  $e_t(\alpha)$ can be also described  in terms of Ruelle transfer operators. For $p\in [1, \infty[$ we set 
$
\|f\|_p^p=\sum_\zeta |f(\zeta)|^p\omega_0(\zeta)$
and define
\begin{equation}
U_p(t)f =f_{-t}{\rm e}^{\frac{t}{p}\Sigma^{-t}}=f_{-t} {\rm e}^{-\frac{1}{p}{\cal S}_{-t}}{\rm e}^{ \frac{1}{p}{\cal S}_0}.
\label{transfer-cl}
\end{equation}
Then  
\[U_p(t_1+t_2)=U_p(t_1)U_p(t_2), \qquad U_p(-t)f U_p(t)g=f_t g, \qquad \|U_p(t) f\|_p=\|f\|_p,\]
i.e., $U_p$ is a group of isometries of the space $L^p({\mathcal E},\omega_0)$ which implements
the dynamics. In terms of this group, one has
\begin{equation}
e_t(\alpha)=\log \|U_{p/\alpha}(t)\textbf{1}\|_p^p,
\label{transfer-ch}
\end{equation}
where $\textbf{1}(\zeta)=1$.

The results described in this section extend under minimal regularity 
assumptions to an essentially arbitrary classical dynamical system \cite{JPR}.
\section{Quantum picture}
Consider  a pair $({\cal K}, H)$ where ${\cal  K}$ is a finite dimensional Hilbert space and $H$ is a Hamiltonian.  Observables are linear maps  $A :{\cal K}\rightarrow {\cal K}$ (the identity map is denoted by $\textbf{1}$) and  states 
are strictly positive density matrices  $\rho$ on ${\cal K}$. The observables evolve in time as $A_t ={\rm e}^{\i t H}A{\rm e}^{-\i t H}$, $t\in {\mathbb R}$, and states as 
$\rho_t={\rm e}^{-\i t H}\rho{\rm e}^{\i t H}$. We write 
$\rho(A)=\tr (A\rho)$.  The relative entropy of two states  $S(\rho, \nu)=\tr(\rho(\log \rho - \log \nu))$ satisfies 
$S(\rho, \nu)\leq 0$ and  $S(\rho, \nu)=0$ iff $\rho=\nu$. 
The R\'enyi relative entropy is defined by $S_\alpha(\rho, \nu)=\log\tr (\rho^{\alpha}\nu^{1-\alpha})$.

Our  starting point is a quantum dynamical system $({\cal K}, H, \omega_0)$, where  $\omega_0$ is a given reference state. 
We assume that $\omega_0$ does not commute with 
$H$ and that the  system is TRI, i.e., that there exists  a complex conjugation on ${\cal K}$ that commutes with $H$ and 
$\omega_0$.

The entropy observable is ${\cal S}_0=-\log\omega_0$. $\Sigma^t = ({\cal S}_t- {\cal S}_0)/t$ is the mean entropy production rate 
observable and   for $t>0$ the finite time second law holds: 
\begin{equation}
\omega_0(\Sigma^t)= -\frac{1}{t}S(\omega_t, \omega_0)\geq 0.
\label{sec-law}
\end{equation}
The entropy production observable (or the quantum phase space contraction rate) is 
$\sigma=-\i [H, \log \omega_0]$ and $\Sigma^t=t^{-1}\int_0^t\sigma_s{\rm d} s$. 

To aid the reader we describe one concrete physical setup. Consider two quantum dynamical systems  $({\cal K}_{l/r}, H_{l/r}, \omega_{l/r})$, colloquially called 
the left and the right. Assume that initially the $l/r$ system is  in  thermal equilibrium at inverse temperature $\beta_{l/r}$, i.e., that 
$\omega_{l/r}=\e^{-\beta_{l/r}H_{l/r}}/Z_{\ell/r}$. Let ${\cal K}={\cal K}_\ell \otimes {\cal K}_{r}$, $\omega_0=\omega_l \otimes\omega_r$, and 
$H= H_l + H_r +V$, where $V$ describes the interaction between the  left and the right system. 
In this case 
\[
\sigma=-\beta_l\Phi_l -\beta_r \Phi_r,
\]
where $\Phi_{l/r}= \i [H_{l/r}, V]$ satisfies 
\[H_{l/r t}- H_{l/r}=-\int_0^t \Phi_{l/r s}{\rm d} s,
\]
and describes the energy flux out of  the $l/r$ system.

We now turn to  quantum  fluctuation relations. TRI implies that the spectrum of $\Sigma^t$ is symmetric w.r.t.\;the origin. 
If $P_{t\lambda}$ is the spectral projection of $\Sigma^t$ onto $\lambda$ and $p^t(\lambda)=\omega_0(P_{\lambda t})$, then the  direct quantization 
of the fluctuation relation (\ref{es-rel}) is 
\begin{equation}
p^t(-\lambda)=\e^{-\lambda t}p^t(\lambda).
\label{q-es-rel-1}
\end{equation}
Setting 
\[e_t(\alpha)=\log \omega_0(\e^{-\alpha t \Sigma^t}),
\label{es-forgot}
\]
one easily verifies that (\ref{q-es-rel-1}) holds iff $e_t(\alpha)=e_t(1-\alpha)$. However, one can show that the quantum Kawasaki identity
$e_t(1)=0$ holds for all $t$ iff $[H, \omega_0]=0$ and so  the direct quantization of  (\ref{es-rel}) fails. The standard 
route of  observable quantization does not lead to quantum  fluctuation relations. 

The following family of quantum entropic functionals indexed by $p\in [1, \infty]$ was introduced in \cite{JOPP}: 
\[
e_{pt}(\alpha)=
\begin{cases}
\log \tr \left(\left[ \e^{-\frac{1-\alpha}{p}{\cal S}_0}\e^{-\frac{2\alpha}{p} {\cal S}_t}\e^{-\frac{1-\alpha}{p}{\cal S}_0}\right]^{\frac{p}{2}}\right)&\text{if $1\leq p<\infty$},\\[3mm]
\log\tr \left(\e^{-(1-\alpha){\cal S}_0 -\alpha{\cal S}_t}\right)&\text{if $p=\infty$}.
\end{cases}
\label{general-func}
\]
To motivate these  functionals, note that 
\[e_{\infty t}(\alpha)=\max_\rho S(\rho, \omega_0)-\alpha t \rho(\Sigma^t),\]
and so $e_{\infty t}(\alpha)$ is the quantization of the variational formula (\ref{var}). We also have 
\[
e_{2t}(\alpha)= S_\alpha(\omega_t, \omega),
\]
and so $e_{2t}(\alpha)$ is the quantization of (\ref{cl-re}). Regarding the other functionals, we need to introduce first the quantization of 
Ruelle's transfer operators. The quantization of the usual $L^p$-norm   is the Araki-Masuda $L^p$-norm 
\[
\|A\|_p^p =\tr \left(\big|A\omega_0^{\frac{1}{p}}\big|^p\right).
\] 
The classical  transfer operators 
(\ref{transfer-cl}) are quantized as 
\[U_p(t)A= A_{-t}\e^{-\frac{1}{p}{\cal S}_{-t}}\e^{\frac{1}{p}{\cal S}_0}.
\]
They satisfy   
\[U_p(t_1+ t_2)=U_p(t_1)U_p(t_2), \qquad U_p(-t)AU_p(t)B=A_tB, \qquad \|U_p(t)A\|_p=\|A\|_p,\]
 and for $p\in [1,\infty[$, 
\[e_{pt}(\alpha)=\log \|U_{p/\alpha}(t)\textbf{1}\|^p_p.
\]
Hence, the functionals $e_{pt}(\alpha)$ for $p\in [1, \infty[$ arise as  the quantization of (\ref{transfer-ch}). 

The functionals $e_{pt}(\alpha)$ have the following properties. The symmetry 
\[
e_{pt}(\alpha)=e_{pt}(1-\alpha),
\]
holds for all $p$ and $\alpha$ and implies the quantum Kawasaki identity
$e_{pt}(1)=e_{pt}(0)=0$. The function $\alpha \rightarrow e_{pt}(\alpha)$ is 
convex and the function $p \mapsto e_{pt}(\alpha)$ is continuous and  decreasing. 
For all $p$, $e_{pt}^\prime(0)=-t\omega_0(\Sigma^t)$. 

The functional $e_{2t}(\alpha)$ has appeared previously in the literature in two seemingly unrelated contexts. 
In \cite{Ku} this functional was related to the fundamental concept of {\em Full Counting 
 Statistics} (FCS) associated to the  repeated measurement protocol of the entropy 
flow. Let ${\cal S}_0=\sum_\lambda \lambda P_\lambda$ 
be the spectral resolution of the entropy observable.  
The probability that  a measurement of ${\cal S}_0$ at time $t=0$ (when the system is  in the state $\omega_0$) yields  $\lambda$ is 
$ \omega_0(P_\lambda)$. After the measurement, the system is in the reduced state 
$\omega_0P_\lambda/{\omega_0(P_\lambda)}$ which evolves in time as 
$\e^{-\i t H}\omega_0P_\lambda \e^{\i t H}/{\omega_0(P_\lambda})$.
Denoting $\lambda'$ the outcome of a second  measurement of ${\cal S}_0$ at a later time $t>0$, 
the joint probability distribution of these two  measurements is $
\tr \left( \e^{-\i t H} \omega_0P_\lambda \e^{\i t H}P_{\lambda^\prime}\right)$. It follows that 
the probability  of observing a mean rate of entropy change $\phi = (\lambda^\prime -\lambda)/t$ is 
\[ {\mathbb P}_t(\phi)= \sum_{\lambda^\prime-\lambda=t\phi}\tr \left( \e^{-\i t H} \omega_0P_\lambda \e^{\i t H}P_{\lambda^\prime}\right). \]
The discrete probability measure ${\mathbb P}_t$ is the FCS for the operationally defined entropy change 
over the time interval $[0, t]$ as specified by the above measurement protocol.  It  is easy to verify  that the  $e_{2t}(\alpha)$ is the 
cumulant generating function for the FCS, i.e.,  
\[ e_{2t}(\alpha)=\log \sum_{\phi}\e^{- t\alpha \phi}{\mathbb P}_t(\phi), \]
and the symmetry $e_{2t}(\alpha)=e_{2t}(1-\alpha)$ yields the fluctuation relation 
\[
{\mathbb P}_t(-\phi)=\e^{-t \phi}{\mathbb P}_t(\phi).
\label{sunday-sick-1}
\]
In \cite{TM} the  functional $e_{2t}(\alpha)$ was motivated by the  algebraic characterization of the Zubarev dynamical ensemble 
\cite{Zu}. Let  ${\cal O}$ be the vector 
space of all linear maps $A: {\cal K}\rightarrow {\cal K}$ equipped with the inner product  $\langle A, B \rangle = \tr (A^\ast B)$. The  relative modular 
operator 
\[ \Delta_{\omega_t|\omega_0}(A)=\omega_t A \omega_0^{-1},
\]
is  a  strictly positive operator on the Hilbert space $({\cal O}, \langle \cdot, \cdot \rangle)$.  Let $Q_t$ be the spectral measure of $-\frac{1}{t}\log \Delta_{\omega_t|\omega_0}$ for the vector $\omega_0^{1/2}\in {\cal O}$. Then 
 \[
 e_{2t}(\alpha)=\log\sum_s \e^{-\alpha t s}Q_t(s),
 \]
and in  particular, ${\mathbb P}_t= Q_t$. This identification provides a striking link between the FCS and modular theory with far reaching 
implications. 
\section{Research program}
In telegraphic terms, our research program can be outlined as follows.\newline
{\bf (a)} The first part of the  program, carried out in \cite{JOP}, deals with development  of the finite time theory  of entropic functionals  in  a general dynamical system setting. 
In the classical case this step is relatively easy \cite{JPR}.  The quantum case is substantially more difficult and the full machinery of Tomita-Takesaki modular 
theory is required. The functionals $e_{pt}(\alpha)$ and quantum Ruelle transfer operators are based on  the  Araki-Masuda theory 
of non-commutative $L^p$-spaces \cite{AM}.   The step {\bf (a)} can be viewed as an abstract unraveling 
of the mathematical structures underlying  the entropic functionals. These structures turn out to be  of considerable conceptual and computational importance. 
\newline
{\bf (b)} This step  concerns the existence and regularity 
properties of the limiting functionals 
\begin{equation}
e_{p+}(\alpha)=\lim_{t\rightarrow \infty}\frac{1}{t}e_{pt}(\alpha),
\end{equation}
and  is a difficult ergodic type problem that   can be studied  only  in the context of concrete models. The existing results 
cover open spin-fermion and spin-boson systems (the proofs are based on the analysis of resonances of quantum Ruelle transfer operators, see also \cite{R} for a pioneering work on the subject), 
and open locally interacting fermionic systems (the proofs are based on $C^\ast$-scattering techniques). In some special cases 
(like the $XY$ chain) the functionals $e_{p+}(\alpha)$ can be expressed in closed form and analyzed in great detail. The step 
{\bf (b)}  of the program has been  carried out in a series of papers and is a joint work with B.~Landon, Y.~Ogata, A.~Panati, Y.~Pautrat,  and M.~Westrich.  It remains 
a  challenge  to extend these results to a wider class of models. We emphasize that the step {\bf (b)} is meaningful only in the context of  infinitely extended models (in other words, 
the thermodynamic limit must precede the large time limit). The thermodynamic limit is also 
needed for the physical interpretation of the   finite time quantum entropic functionals of infinitely extended systems.  \newline
{\bf (c)} The Legendre transform of the  limiting functional $e_{2+}(\alpha)$  is the rate function describing  the large deviation fluctuations 
 of the full counting statistics  as $t\rightarrow \infty$. In the linear regime and under suitable regularity assumptions 
 the existence of $e_{2+}(\alpha)$ also implies  the central limit theorem 
 for the full counting statistics. For open quantum systems and in  the linear regime (near equilibrium and for small $\alpha$) all  functionals $e_{p+}(\alpha)$ reduce to 
 Green-Kubo formulas for  heat/charge currents.  However, the  development  of 
  linear response theory that  goes  beyond open quantum systems and allows   for general thermodynamical/mechanical forces can be based 
  only on the functional $e_{\infty+}(\alpha)$ (see \cite{G, JPR} for the results in the 
 classical case that motivated this development). The functionals $e_{p+}(\alpha)$, together 
 with symmetries $e_{p+}(\alpha)=e_{p+}(1-\alpha)$,  can 
 be viewed as extensions of the fluctuation-dissipation theorem to the far from equilibrium regime. \newline
 {\bf (d)} The functional $e_{2+}(\alpha)$ coincides with the  Chernoff error 
 exponent for the quantum hypothesis testing of the arrow of time and this  links quantum hypothesis testing, a rapidly 
 developing branch of quantum information theory,  to  non-equilibrium statistical mechanics. This connection has been  explored in \cite{JOPS}.   \newline
 {\bf (e)} We have only discussed the entropic functionals defined with respect to the reference (initial) state of the system (they 
 are sometimes called Evans-Searles type functionals). The Gallavotti-Cohen type entropic functionals are defined with respect to the non-equilibrium steady state (NESS) (the state that the  infinitely extended system reaches in the large time limit). The Gallavotti-Cohen 
 type functionals  are considerably more technical 
 to introduce and study, and in the quantum case have a somewhat delicate physical interpretation. The {\em Principle of Regular Entropic Fluctuations} (PREF) introduced in \cite{JPR, JOP} asserts that under normal conditions  the limiting entropic functional of the Evans-Searles  and Gallavoti-Cohen type are identical. Since  under normal conditions  the NESS is singular with respect to the reference state,  the PREF can be viewed as a strong ergodic property of the physical model under consideration.  \newline
{\bf (f)} The developement of nonequilibrium statistical mechanics of open quantum systems in the Markovian approximation started with the pioneering work of Lebowitz and Spohn \cite{LS}.  Since
the Markovian approximation is often the only technically accessible way to describe an open system,
it is also important to develop a general theory of entropic fluctuations in this framework. The first
attempts in this direction are due to Derezi\'nski, De Roeck and Maes \cite{RM,DRM}. In \cite{JPW},
we derive fluctuation relations starting from structural properties of the generator (Lindbladian) of
the Markovian dynamics.

\section{Remarks} 
Non-equilibrium statistical mechanics is a difficult subject and for many years our theoretical understanding has been restricted to  
linear regime near equilibrium (linear response theory, the fluctuation-dissipation theorem). There are good reasons for this: the richness and variety 
of non-equilibrium phenomena  indicate that far from equilibrium physics may tolerate 
very few universal constraints. The recently discovered fluctuation relations of Evans-Searles and Gallavotti-Cohen  are two universally valid constraints that hold 
far from equilibrium and reduce to linear response  near equilibrium. These insights and subsequent developments (see \cite{JPR}
for references) have dramatically altered our understanding of classical non-equilibrium statistical mechanics. The extensions of fluctuation 
relations to quantum domain have led to further surprises  that still remain to be fully explored. The research program outlined in this note 
is a first step in this direction. 


\end{document}